# Evaluation methodology of Model Predictive Controllers for building's energy systems

## Ali CHOUMAN*[1, 2], Peter RIEDERER [1], Frédéric WURTZ [2]

**[1] Centre Scientifique et Technique du Bâtiment (CSTB),
F-06904 Sophia Antipolis, France**

**[2] Univ. Grenoble Alpes, CNRS, Grenoble INP, G2Elab,
F-38000 Grenoble, France**

***ali.chouman@cstb.fr**

*ABSTRACT. Climate change poses a serious threat to the Earth's ecosystems, fueled primarily by escalating greenhouse gas emissions. Among the main contributors, the building sector stands out due to its significant energy demand. Addressing this challenge requires innovative techniques in the control of energy systems in buildings. This paper deals with the formulation of a methodology designed to evaluate the performance of these controllers. The evaluation process involves the establishment of a comprehensive test protocol and a diverse set of scenarios to evaluate the controllers. Key performance indicators are used to quantify their effectiveness based on the test results. A practical case study is presented as an application to introduce this methodology, focusing on the integration of Model Predictive Controllers (MPCs) with the Dimosim thermal simulation platform. The digital twin of the Greener building in Grenoble is used as a model for emulation. The paper demonstrates the ability of the proposed methodology to test and rank MPCs in different test scenarios, providing valuable feedback on their performance capabilities. The paper highlights the importance of the developed approach in systematically evaluating and ranking MPCs for optimized building energy management.*

## 1. INTRODUCTION

One of the most critical challenges facing society today is climate change, which underscores the urgent need for substantial energy savings, particularly in terms of fossil resources. Global concern regarding the environmental impact of energy consumption has significantly grown in recent years. And since buildings account for about 38% of global final energy use (GSR for Buildings and Construction 2020), energy-efficient building control can have an important contribution. To enhance energy efficiency and adaptability in building energy systems control, advanced control devices have emerged, incorporating techniques like AI (artificial intelligence) and machine learning (Schreiber et al. 2021). These techniques lead towards the optimal management of energy systems aiming to optimize energy consumption, thermal comfort, and energy flexibility. Among the innovative energy management modules, our focus will be more on the predictive innovative control algorithm, or the MPC (Model predictive Controllers). This type of device or control algorithm aims to optimize energy management and meet the specific thermal requirements of building zones by accurately predicting and selecting optimal control strategies. In this article, we introduce a methodology for evaluating the performance of energy control systems in buildings. Through a comprehensive case study involving the coupling of control algorithms with a dynamic simulator, we utilize performance indicators to quantify the benefits of employing these algorithms and establish the first step of a ranking system for comparative analysis.

## 2. STATE OF ART FOR EXISTING EVALUATION METHODOLOGIES

Many reviews such as the one done by (Afram et Janabi-Sharifi 2014) compare the classical control approaches to MPC in different applications. The classical control examples shown in this study consist of the most commonly used control techniques, such as on/off control and P, PI, and PID control (Process Controllers - P, PI & PID, s. d.). In this comparative literature study, the authors expose several use cases of classical and innovative control approaches. Their findings robustly support the superiority of the MPC approach in numerous applications.

Nonetheless, translating this potential into quantifiable and comparable benefits remains a complex task. In most cases, the evaluation is done according to a predefined scenario of the test. In many cases, researchers gauge performance through metrics for a specific case of test, such as the percentage variance in energy consumption as in the study realized by (Goyal, Ingley, et Barooah 2012), or by quantifying the optimized operation costs and some thermal comfort indicators such as predicted percentage of dissatisfied (PPD) as in the study of (Huang et al. 2021).

Considering our specific focus on methodologies employed for MPC evaluation, this section provides an overview of some existing research in this domain. It is worth noting that there's a relative scarcity of research dedicated to the development of comprehensive evaluation methods. Our ultimate objective is to propose a robust methodology for assessing MPC's true potential, factoring in real-world complexities and variables, in the form of a standardized, reference evaluation method. One of the proposed methodologies in the literature is the Building Optimisation Testing Framework (BOPTEST), established by (Blum et al. 2021) and focused on the benchmarking of control strategies in buildings. This framework presents an interesting highlight of the evaluation concepts used in the case studies of the advanced control systems in buildings' HVAC implementation. According to this review, the existing literature evaluates advanced controllers individually through case studies that differ in building types, evaluation metrics, and comparative benchmarks. The framework in BOPTEST offers a standardized set of test cases, including building emulators, a Run-Time Environment (RTE) where the controller is coupled to the emulator, and common calculations of key performance indicators (KPIs). This facilitates the testing and evaluation of advanced controllers. One more interesting work that we will refer to is the methodology established by (Huang et al. 2021). This paper proposes a protocol used to evaluate the performance of MPC over different choices of modeling and control parameters and provides a simulation routine to achieve this evaluation. The evaluation routine proposed in this paper is based on simulation and it is automated and consists of four consecutive steps. It begins by generating the test scenario of MPC parameters, then a model identification, a control implementation of the MPC, and in the last step, metric evaluation.

Following this review of the relevant work for MPC evaluation, it is observed that a significant challenge lies in the heavy reliance on a set of predefined or case-specific tests for performance evaluation. This approach might not fully capture the controller's effectiveness across diverse and unpredictable operational scenarios. We will present in the next section our proposed approach for the evaluation followed by a case study involving the coupling of control algorithms with a dynamic simulator. Using this methodology, we propose a protocol and a set of tests for evaluation and a case study application for some scenarios of test to demonstrate the results of benchmarking two MPCs and a reactive controller.

## 3. METHODOLOGY FOR MPC EVALUATION

### 3.1 SPECIFICATIONS OF THE NEW METHODOLOGY

The evaluation of controllers for building energy systems requires a global approach that considers their operating principles, adaptability, and operation scenario. Key aspects include evaluating the learning period of the controllers, especially for algorithms with historical data training models, the prediction horizon for MPC algorithms, their performance in different building types, climates, and internal conditions, and their robustness to uncertainties. Performance metrics such as energy savings, occupant comfort, and cost-effectiveness are critical, as is an examination of user interaction and the potential for manual overrides. This comprehensive evaluation ensures that controllers are effectively evaluated for their ability to optimize building energy use, maintain occupant comfort, and adapt to changing environmental and operational conditions.

### 3.2 FORMULATION OF THE EVALUATION METHODOLOGY

To effectively evaluate building energy system controllers, a methodology integrating the controller with either a physical building or a detailed thermal simulation model of a building is essential. First, the methodology involves choosing a suitable building model for testing and establishing a baseline scenario for the test. Key steps include outlining the testing protocols, specifying data requirements for predictive controllers, and detailing the analysis approach for interpreting results. Objectives and key performance indicators (KPIs) like energy efficiency and comfort must be defined. In the last step, the evaluation methodology ends with the creation of a comprehensive test package that includes various scenarios capable of simulating the possible internal and external environmental conditions, building characteristics, and potential disturbances. This approach ensures that the robustness of controller performance is thoroughly evaluated against a baseline scenario and in varied conditions. In addition, a critical validation step confirms the accuracy of the evaluation outcomes. Through this methodology, controllers can be effectively ranked and compared on a multi-criteria basis, providing valuable conclusions about their efficacy.

### 3.3 TEST PROTOCOL AND SCENARIOS

The proposed evaluation process of control algorithms involves several steps as shown in Figure 1:

- Select the thermal zone where the control algorithm will be implemented, which could be either a real building or a numerical thermal model, as in our case study;

- Determine the scenario of the test by selecting the periods of training and operation, the duration of the test, and the horizon of predictions, in case of using a predictive controller. This step ensures that the test conditions are accurately defined and consistent across all tested algorithms;

- Select the control algorithm to be coupled with the thermal model. In our case study, we have the two developed MPC algorithms and a reactive controller for illustration purposes;

- Couple the algorithms to the model where the algorithm will have complete control over the energy systems of the thermal model during the operation duration, by imposing the set-point temperature on those systems.

- Once the coupling process is completed, a simulation is launched, and the algorithm takes control of the energy systems of the model for the test duration;

- After the simulation is completed, the results necessary for the key performance indicators (KPIs) calculations are collected and used to determine the numerical values of the indicators;

- The final step is to rank the algorithms based on their performance, a radar chart plot is used to visualize the field of domination of each controller among the evaluation indicators. This method ensures that the evaluation process is objective (no ponderation used for the indicators) and can be easily understood. The results obtained from the evaluation of multiple algorithms are used to rank the controllers based on their scores for each KPI.

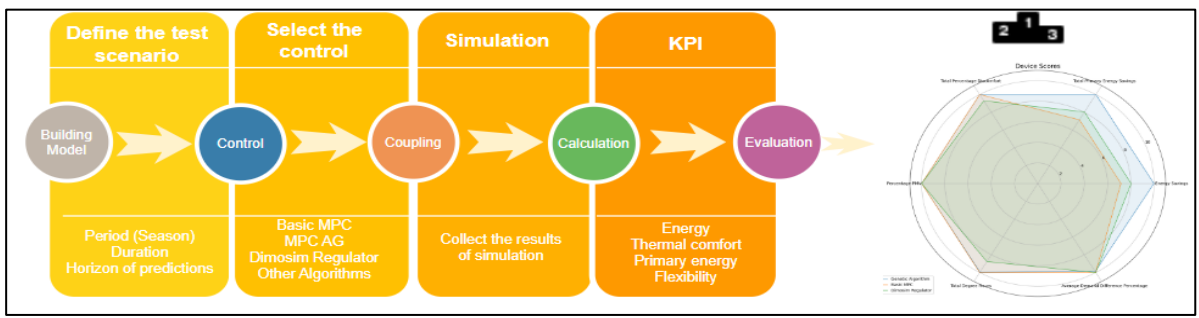

*Figure 1: A flowchart for the protocol proposed in the methodology of evaluation.*

The described protocol will be systematically applied across a range of crucial scenarios for validating the effectiveness of the controllers. This approach involves a comprehensive examination of the algorithms across a diverse set of conditions. A battery of possible scenarios of testing is defined for this purpose. The scenarios consider different internal conditions (internal gains and occupancy profiles), external conditions (weather scenarios), and building characteristics (envelope, usage type, etc.), besides different possible disturbances and occupants' actions. This testing process aims to ensure that the algorithms are not only effective in a single, specific scenario, but are also robust under a wider range of different conditions. And thus, ensuring that our approach applies generally and isn't limited to specific situations or scenarios.

Depending on the control variable being controlled by the algorithm of control (air temperature, air quality, DHW, etc.), our methodology outlines various scenarios that affect these control variables. In this section, we will present examples of scenarios tailored to room temperature control within a specific zone. These examples will serve as the basis for further testing of algorithmic aimed at optimizing indoor air temperature, as will be presented in the following case study section.

The scenarios outlined are designed to test temperature control algorithms within buildings, focusing on how different variables influence energy systems' performance and indoor environment quality. The scenarios include two categories of variables: static and dynamic. Static variables include characteristics that do not change over time, such as the building's envelope U-value (both opaque and transparent elements), the percentage of external facades, building inertia, floor position, zone orientation, window-to-wall ratio (for solar gains scenarios), insulation positioning, building type, and heating or cooling system efficiency. On the other hand, dynamic variables are subject to change and are mostly stochastic. These include occupancy profiles, internal gain profiles, operation of openings (doors and windows), use of blinds (whether manually controlled or left to occupant discretion), occupant-modified settings (temperature, humidity, CO2 levels), uncontrolled ventilation rates, and weather conditions.

By testing control algorithms against combinations of static and dynamic variables, we are considering both the predictable aspects of building performance and the unpredictable nature of human behavior and environmental conditions. Thus, this approach identifies a wide range of conditions to test the algorithms.

### 3.4 Key performance indicators (KPIs)

Key Performance Indicators (KPIs) are a set of measurable parameters that are used to evaluate the performance of a control algorithm or product regarding the results of its implementation. The selection of KPIs for a specific MPC is based on the objectives and requirements of the energy management system. Therefore, KPIs play a crucial role in evaluating the performance of the MPC by providing quantitative measurements of its effectiveness in meeting the desired objectives. These KPIs can be further broken down into sub-KPIs such as peak demand, setpoint deviations, and temperature fluctuations. The KPIs defined in this methodology are derived from the European project Collectief (COLLECTiEF 2021), which leads to implementing innovative control technologies in buildings to improve building energy performance and contribute to global climate and energy goals. The selected KPIs can be classified into four main categories, with some sub-KPIs for each category. The four main categories of KPIs are energy savings, primary energy savings, comfort analysis, and flexibility analysis. For each KPI, a set of sub-KPIs are defined such as the percentage of energy saved, cost savings, percentage of time outside the comfort temperature range, peak power reduction during demand response event, etc. Other qualitative KPIs could be also used such as the amount of historical data required to train the optimization model in the algorithm, and the time needed for computation.

## 4. Case study : an introduction to the use of defined methodology in algorithms evaluation

This section's case study showcases the ability of our developed protocol to evaluate and comparatively rank various controllers through a multi-criteria approach. The case study presents a baseline evaluation scenario for the controllers under review, leading to an initial ranking presented in the form of a radar chart. It's important to note that the rankings displayed are preliminary and will be validated against a series of defined scenarios in subsequent studies, which will lead to the final evaluation of the algorithms. This approach underscores the importance of robustness in controller performance across varying operational conditions.

### 4.1 Building model

The selection of a case-study-building model is crucial in ensuring that the results of the study are relevant and accurate. In this context, the Predis-MHI zone of the Electrical Engineering Lab at Grenoble University (G2ELab) was chosen due to its suitability as a testbed for evaluating different control strategies. This studied zone is a living lab and it is situated on the fourth floor of the Green'ER building, which is located in Grenoble, France (Delinchant et al. 2016). In a related article, we presented the procedure adopted to create and calibrate a thermal model of the building, using the interesting amount of data and measures available due to its monitoring system. The calibration ensures that the model reflects the thermal behavior of the building. The simulation would be ruled by Dimosim (Garreau et al. 2021), which is a bottom-up dynamic simulation platform for energy systems in buildings and districts developed since 2013 by the CSTB (www.cstb.fr). This tool is built on an object-oriented structure.

### 4.2 CONTROL ALGORITHMS

The case study involves developing MPC algorithms that produce a control strategy leading to minimizing energy consumption while maintaining thermal comfort levels in a given zone. The algorithms are coupled to Dimosim emulator to test their performance. The outdoor temperature and solar gains that can affect the temperature evolution in the zone are anticipated by the algorithms within the prediction horizon, offering the algorithms the flexibility to pre-react regarding the predicted variable impact, while satisfying optimization constraints. Anticipated occupant profiles are also considered to prevent thermal discomfort during the operation. The algorithms were developed to optimize the value of the temperature set point and manage the energy systems of the zone. The interaction between the algorithm and the emulator is bi-directional, with the controller receiving the initial states of the thermal zone from the emulator at the beginning of each prediction horizon, and sending the control strategy after data treatment.

In our case study, we developed two MPC algorithms that will be compared to the reactive controller already implemented in Dimosim simulation platform. For developed algorithms, the first is a basic predictive combinatorial algorithm. The algorithm predicts the internal and external conditions in the thermal zone for a horizon of time. Then, it proceeds to test various combinations of temperature set-point vectors over the predicted time horizon. The algorithm identifies after that the optimal solution that minimizes energy consumption without compromising thermal comfort within the thermal zone. The second controller is based on a genetic algorithm to overcome the limitations of the combinatorial approach in generating and testing control possibilities. This algorithm uses a genetic algorithm to minimize an objective function as the optimization is described as a mathematical problem. The optimization problem involves minimizing an objective function that takes into account both energy consumption and thermal comfort while dealing with flexibility signals. To achieve this optimization, the algorithm perfectly predicts the state variables of the controlled system, thus, the algorithm has full access to the emulator's data, ensuring that the MPC has complete knowledge of the parameters and internal conditions in the emulator, to which it is expected to provide a control strategy.

### 4.3 SCENARIO OF TEST

For this first application of the methodology, the baseline scenario used for the test is defined as follows, a week of operation in the heating season. Real weather data are used for the test, and the baseline model of the building is coupled with the MPCS. For both developed MPCs, the horizon for predictions is set to be 6 hours with hourly time step for operation. This scenario is a baseline scenario and its results of evaluation will be validated using a set of scenarios including different weather, building envelopes, and operation periods scenarios, to test the robustness of the controller with the variation of the test conditions.

### 4.4 RESULTS AND DISCUSSION

As shown in the proposed methodology, after choosing the model and the scenario test, besides the algorithms to be tested and ranked, each algorithm is coupled with the DIMOSIM emulator to assess its performance. The results of the baseline scenario are shown in Figure 2.

The results of both algorithms are used to calculate performance indicators based on optimization objectives to quantify the gains of their implementation. The results are compared to the results of the reactive regulator implemented in the emulator (Dimosim), which serves to maintain the temperature around its set point, without any way to predict future behavior. Our method involves rating KPIs on a scale of 0 to 10 in the radar chart. The highest performance achieved is set at 10, and all other values are normalized relative to this top score, representing their scores between 0 and 10. The results of the baseline scenario underline that the developed Model Predictive Controllers (MPCs) outperform the reactive control in terms of both energy savings and comfort, in addition to their responsiveness to flexibility signals, except for Total Degree Hours sub-KPI which represent the time and magnitude of temperature deviation outside of the comfort range (°C.hour). The importance of these results lies in the ability of the introduced protocol to evaluate and rank several controllers under the same conditions using a multi-criteria approach. It's important to note that the ranking provided in this study is primarily intended to demonstrate the effectiveness of the method in evaluating and classifying controllers. Future work will involve testing these controllers in different scenarios, as previously discussed, to accurately confirm their rankings. This upcoming phase will focus on creating a comprehensive set of scenarios that include a wide range of profiles and variables for the validation of calculated performance.

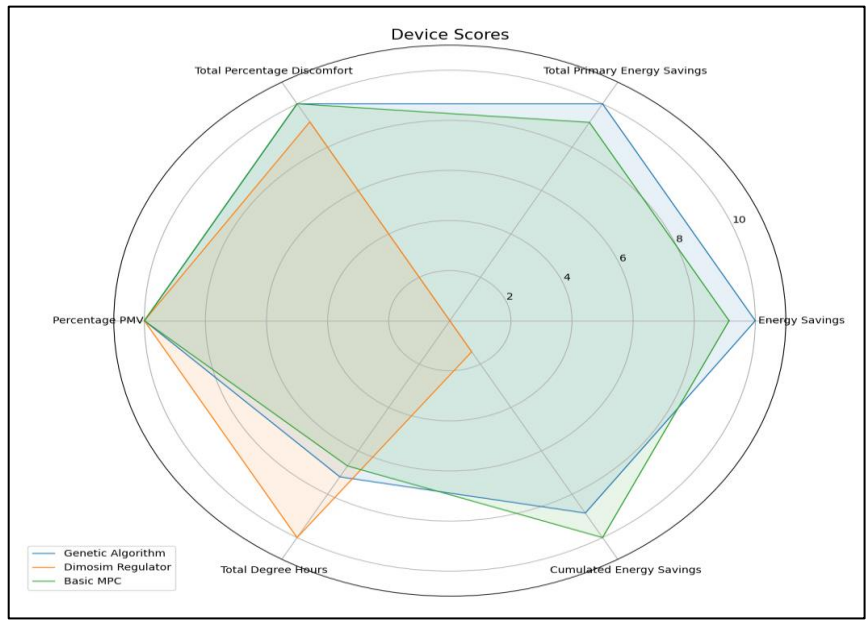

*Figure 2:Radar chart shows the scores of the evaluated devices according to some KPI for the baseline scenario.*

## 5. FUTURE WORK

As mentioned before, the next phase is to validate the results of the baseline scenario by conducting further testing with the various scenario combinations outlined in Section 3.3. Given the large number of possible scenario combinations, a critical future direction is to optimize the testing process. This involves conducting a detailed analysis to identify the parameters that have the most significant impact on the scenarios. The goal is to identify a refined set of scenarios to test, optimizing the validation process for the calculated performance metrics. This approach not only ensures the robustness of the performance evaluation but also increases the efficiency of the methodology by focusing on the most influential variables. Future work includes also introducing additional methods for ranking and analysis such as Pareto graphs to see the interaction between the KPIs when evaluating the performance of controllers.

## 6. CONCLUSION

This study presents an introduction to a methodology for evaluating innovative energy management systems and control strategies in building environments, a case study is presented also to show the adopted protocol application mechanism for a baseline scenario. The primary objective is to establish a robust evaluation protocol and a comprehensive set of tests. These measures are critical to accurately assess performance levels and facilitate multi-criteria comparison to effectively rank different control systems. The variety of tests in the proposed method prevents that control algorithms may be tuned to detect and perform optimally only under certain test conditions, and thus may not reflect true operational performance(analogy to the "Dieselgate" scandal (Palmer et Schwanen 2019). Our methodology is designed to ensure a transparent and fair assessment by covering a wide range of operational scenarios.